\documentclass[%
 reprint,
 showkeys,
superscriptaddress,
 amsmath,amssymb,
 aps,
]{revtex4-2}

\usepackage{graphicx}%
\usepackage{amsmath,amssymb,amsfonts}%
\usepackage{tensor}

\usepackage{mathrsfs}%
\usepackage{listings}%

\usepackage{comment}%

\usepackage{tabularx}
\usepackage{multirow}
\usepackage{booktabs}
\usepackage{array}
\newcolumntype{Y}{>{\raggedright\arraybackslash}X}
\usepackage[dvipsnames,table]{xcolor}
\usepackage[breaklinks,colorlinks, urlcolor=Cerulean,citecolor=RedViolet,linkcolor=MidnightBlue]{hyperref}
\usepackage{orcidlink}

\usepackage{dcolumn}
\usepackage{bm}%
\usepackage[mathlines]{lineno}%

\newcommand{\ve}{\varepsilon}

\newcommand{\T}{\tensor}

\newcommand{\cE}{\mathcal E}

\newcommand{\cU}{\mathcal U}

\newcommand{\ha}{\hat \alpha}
\newcommand{\hb}{\hat \beta}

\newcommand{\hs}{\hat \sigma}

\newcommand{\hk}{\hat \kappa}
\newcommand{\hmu}{\hat \mu}
\newcommand{\hn}{\hat \nu}

\DeclareMathOperator{\Tr}{Tr}

\makeatletter
\let\jnl@style\rmfamily    %
\makeatother

\def\aj{\ref@jnl{AJ}}                   %
\def\actaa{\ref@jnl{Acta Astron.}}      %
\def\araa{\ref@jnl{ARA\&A}}             %
\def\apj{\ref@jnl{ApJ}}                 %
\def\apjl{\ref@jnl{ApJ}}                %
\def\apjs{\ref@jnl{ApJS}}               %
\def\ao{\ref@jnl{Appl.~Opt.}}           %
\def\apss{\ref@jnl{Ap\&SS}}             %
\def\aap{\ref@jnl{A\&A}}                %
\def\aapr{\ref@jnl{A\&A~Rev.}}          %
\def\aaps{\ref@jnl{A\&AS}}              %
\def\azh{\ref@jnl{AZh}}                 %
\def\baas{\ref@jnl{BAAS}}               %
\def\bac{\ref@jnl{Bull. astr. Inst. Czechosl.}}
\def\caa{\ref@jnl{Chinese Astron. Astrophys.}}
\def\cjaa{\ref@jnl{Chinese J. Astron. Astrophys.}}
\def\icarus{\ref@jnl{Icarus}}           %
\def\jcap{\ref@jnl{J. Cosmology Astropart. Phys.}}
\def\jrasc{\ref@jnl{JRASC}}             %
\def\memras{\ref@jnl{MmRAS}}            %
\def\mnras{\ref@jnl{MNRAS}}             %
\def\na{\ref@jnl{New A}}                %
\def\nar{\ref@jnl{New A Rev.}}          %
\def\pra{\ref@jnl{Phys.~Rev.~A}}        %
\def\prb{\ref@jnl{Phys.~Rev.~B}}        %
\def\prc{\ref@jnl{Phys.~Rev.~C}}        %
\def\prd{\ref@jnl{Phys.~Rev.~D}}        %
\def\pre{\ref@jnl{Phys.~Rev.~E}}        %
\def\prl{\ref@jnl{Phys.~Rev.~Lett.}}    %
\def\pasa{\ref@jnl{PASA}}               %
\def\pasp{\ref@jnl{PASP}}               %
\def\pasj{\ref@jnl{PASJ}}               %
\def\rmxaa{\ref@jnl{Rev. Mexicana Astron. Astrofis.}}%
\def\qjras{\ref@jnl{QJRAS}}             %
\def\skytel{\ref@jnl{S\&T}}             %
\def\solphys{\ref@jnl{Sol.~Phys.}}      %
\def\sovast{\ref@jnl{Soviet~Ast.}}      %
\def\ssr{\ref@jnl{Space~Sci.~Rev.}}     %
\def\zap{\ref@jnl{ZAp}}                 %
\def\nat{\ref@jnl{Nature}}              %
\def\iaucirc{\ref@jnl{IAU~Circ.}}       %
\def\aplett{\ref@jnl{Astrophys.~Lett.}} %
\def\apspr{\ref@jnl{Astrophys.~Space~Phys.~Res.}}
\def\bain{\ref@jnl{Bull.~Astron.~Inst.~Netherlands}}
\def\fcp{\ref@jnl{Fund.~Cosmic~Phys.}}  %
\def\gca{\ref@jnl{Geochim.~Cosmochim.~Acta}}   %
\def\grl{\ref@jnl{Geophys.~Res.~Lett.}} %
\def\jcp{\ref@jnl{J.~Chem.~Phys.}}      %
\def\jgr{\ref@jnl{J.~Geophys.~Res.}}    %
\def\jqsrt{\ref@jnl{J.~Quant.~Spec.~Radiat.~Transf.}}
\def\memsai{\ref@jnl{Mem.~Soc.~Astron.~Italiana}}
\def\nphysa{\ref@jnl{Nucl.~Phys.~A}}   %
\def\physrep{\ref@jnl{Phys.~Rep.}}   %
\def\physscr{\ref@jnl{Phys.~Scr}}   %
\def\planss{\ref@jnl{Planet.~Space~Sci.}}   %
\def\procspie{\ref@jnl{Proc.~SPIE}}   %

\begin{document}

\title{Toward a Theory of Gravitational Wave Turbulence}

\author{Holly Krynicki\,\orcidlink{0000-0002-7891-2515}} 
\affiliation{Department of Mathematics, California Institute of Technology, Pasadena, CA 91125, USA}
\affiliation{TAPIR, Mailcode 350-17, California Institute of Technology, Pasadena, CA 91125, USA}

\author{Jiaxi Wu\,\orcidlink{0000-0003-3829-967X}} 
\affiliation{TAPIR, Mailcode 350-17, California Institute of Technology, Pasadena, CA 91125, USA}

\author{Elias R. Most\,\orcidlink{0000-0002-0491-1210}} 
\affiliation{TAPIR, Mailcode 350-17, California Institute of Technology, Pasadena, CA 91125, USA}
\affiliation{Walter Burke Institute for Theoretical Physics, California Institute of Technology, Pasadena, CA 91125, USA}

\begin{abstract}
 General relativity describes the dynamics of gravitational waves, which can feature nonlinear interactions, such as those underlying turbulent processes. Theoretical and numerical explorations have demonstrated the existence of gravitational wave turbulence, of which a full and general mathematical description is currently not known. Here, we take essential steps towards such a theory.
 Leveraging a formulation exactly recasting general relativity as a set of nonlinear electrodynamics equations, we demonstrate that general relativity 
 admits an Elsasser formulation---the same type of equation underpinning magnetohydrodynamic turbulence.
 We further show that nonlinear interactions described by this equation are in part Alfv\'enic, linking gravitational wave turbulence to Alfv\'enic turbulence.
 Our work paves the way for a new understanding of nonlinear gravitational wave dynamics through insights from magnetohydrodynamics.
\end{abstract}

\maketitle

\section{Introduction}

Gravitational waves are a fundamental prediction of the theory of general relativity \cite{1916SPAW.......688E}, and their existence has been confirmed indirectly \cite{1989ApJ...345..434T} and, most recently, directly \cite{LIGOScientific:2016aoc} with the detection of hundreds of gravitational wave signals from merging compact objects \cite{LIGOScientific:2025slb}.
The detection of the ringdown part of the gravitational wave signal \cite{Giesler:2019uxc} (including a recent high signal-to-noise event \cite{KAGRA:2025oiz}) has sparked considerable interest in nonlinear aspects of gravitational wave and black hole dynamics. This includes imprints of quasi-normal modes onto the gravitational wave signal \cite{LIGOScientific:2016lio}, at which point black hole ringdown enters a linear regime with overtones \cite{Giesler:2019uxc,Carullo:2019flw,Cotesta:2022pci,Finch:2022ynt,Baibhav:2023clw,Giesler:2024hcr,Cheung:2023vki}, and, in particular for future detections, the presence of gravitational wave memory (see, e.g., Ref. \cite{Mitman:2024uss} for a recent review) and nonlinear couplings of quasi-normal modes \cite{Sberna:2021eui,Mitman:2022qdl,Cheung:2022rbm}. 
From a theoretical standpoint, since gravitational wave interactions are fundamentally nonlinear, they can give rise to exotic phenomena, such as black hole formation \cite{Pretorius:2018lfb} in the strongly nonlinear regime and turbulence for weaker interactions \cite{Galtier:2017mve,Galtier:2021ovg,Gay:2025unv}. In particular, the latter point has garnered considerable interest over the past years. Motivated by the fluid--gravity duality for asymptotically anti-de Sitter spacetimes \cite{Bhattacharyya:2007vjd}, several works have investigated the presence of turbulence, either by solving two-dimensional conformal relativistic fluid dynamics on the fluid side \cite{Carrasco:2012nf,Adams:2013vsa} or by studying black hole perturbations on the gravity side \cite{Green:2013zba}, finding the presence of a parametric instability reminiscent of the onset of a turbulent cascade \cite{Yang:2014tla}.
Enabled by progress on direct numerical solutions of the Einstein equations, driven gravitational wave turbulence and the emergence of an inverse cascade have recently been demonstrated \cite{Ma:2025rnv}.

Despite this progress, a full and general theoretical description of gravitational wave turbulence is currently missing. Under the assumptions of strong symmetries in the underlying metric, Ref. \cite{Galtier:2017mve} has investigated weak turbulence using a wave turbulence approach. Some of the properties of gravitational wave turbulence they found in this regime bear surprising similarities to magnetohydrodynamic (MHD) turbulence (see, e.g., Ref. \cite{Schekochihin:2020aqu} for a recent review).
In a MHD description, turbulence arises through a nonlinear interaction of counter-propagating Alfv\'en waves \cite{1964SvA.....7..566I,1965PhFl....8.1385K,1995ApJ...438..763G}. In what follows, we depict how these may be seen. Focusing on the case of ideal relativistic MHD, given a spacetime metric \(g^{\mu \nu}\) and an energy--momentum tensor \cite{Chandran:2017zdg}
\begin{align}\label{eqn:tmunu_MHD}
    T^{\mu \nu}_{\rm MHD} = \mathcal{E} u^\mu u^\nu + {\mathcal{E}\Pi} g^{\mu \nu} -b^\mu b^\nu\,,
\end{align}
where $u^\mu$ the fluid velocity, $b^\mu$ the magnetic field seen in the frame comoving with the fluid, $\mathcal{E} = e + P + b^2$, $e$ and $P$ being fluid energy density and pressure, and ${\Pi} = (P + b^2/2)/\mathcal{E}$.
One of the fundamental insights {from MHD} is that the equation of motion, $\nabla_\mu T^{\mu\nu}_{\rm MHD}=0$, can be written in a form that manifestly demonstrates the nonlinear wave nature. In this Elsasser form \cite{Elsasser:1950zz}, the relativistic MHD equations read \cite{Chandran:2017zdg,TenBarge:2021qmk}:
\begin{align}\label{eqn:elsasser_mhd}
   \nabla_\nu\left[ z_{\pm}^\mu z^{\nu}_\mp\right. &+ \left.\Pi g^{\mu \nu}\right]\nonumber \\
   &+ \left(\frac{3}{4} z_{\pm}^\mu z^{\nu}_\mp +\frac{1}{4} z_{\mp}^\mu z^{\nu}_\pm + \Pi g^{\mu \nu} \right) \partial_\nu\log \mathcal{E} =0\,.
\end{align}
This equation is the starting point for many of, if not most of, MHD turbulence analyses \cite{Schekochihin:2020aqu}.
The theory primarily describes a set of Elsasser variables, namely
\begin{align}\label{eqn:z_mhd}
    z_\pm^\mu = u^\mu \pm \frac{b^\mu}{\sqrt{\mathcal{E}}}\,.
\end{align}
These variables incorporate the propagation of Alfv\'en waves (transverse waves along the magnetic field), which can interact nonlinearly only when counter-propagating (e.g., Ref. \cite{1987JGR....92.7363M}). The nonlinear interaction of these Alfv\'en waves then creates a turbulent cascade, which is intrinsically anisotropic relative to the magnetic field, especially in the critically balanced strong turbulence regime \cite{1995ApJ...438..763G}.

While at first glance {MHD and general relativity} are seemingly disjoint from the question of gravitational wave turbulence, we demonstrate that these two fields are intimately linked due to the underlying electrodynamic structure of spacetime \cite{Maartens:1997fg,Nichols:2012jn,Owen_2011}.
Mathematically, general relativity can be shown to be exactly equivalent to a system of nonlinear electrodynamics equations \cite{Olivares_2022,Peshkov:2022cbi,Boyeneni:2025tsx}. Indeed, the resulting electric and magnetic field dynamics close to mergers \cite{Boyeneni:2025tsx} bear some resemblance to force-free electrodynamics behavior \cite{Most:2020ami,Most:2023unc,Most:2024qgc}, which features turbulence \cite{TenBarge:2021qmk,Ripperda:2021pzt}.

Exploiting this exact analogy, we demonstrate herein that under conditions relevant to gravitational wave turbulence, the effective stress--energy (pseudo-)tensor of general relativity takes the form of Eq. \eqref{eqn:tmunu_MHD} in the force-free electrodynamic limit \cite{TenBarge:2021qmk,Ripperda:2021pzt}. Next, we illustrate that general relativity possesses an Elsasser equation akin to Eq. \eqref{eqn:elsasser_mhd} with tensor Elsasser variables, which encode perturbations of the gravitational magnetic field relative to a freely-falling observer. Finally, we show that nonlinear interactions of these perturbations contain fundamentally Alfv\'enic pieces; i.e., they involve a coupling of opposite Elsasser variables.

In  this work, we use units $G=c=1$ and denote 
$\ve ^{abcd} = \frac{-1}{\sqrt{-g}} \epsilon ^{abcd}\,,~ \ve _{abcd} = \sqrt{-g} \epsilon _{abcd}$
as the Levi-Civita tensor and tensor density, respectively. Additionally, we denote the spacetime metric with $g_{\mu\nu}$ and its determinant with $g$.

\section{General relativity as nonlinear electrodynamics}\label{sec:einstein}

In order to leverage insights from MHD turbulence theory in a gravitational context, we need to express general relativity in a form most closely resembling (nonlinear) electrodynamics. 
For this purpose, we use the DGREM formulation of Ref. \cite{Olivares_2022,Peshkov:2022cbi} (see also Refs. \cite{Boyeneni:2025tsx,Estabrook:1996wa,Buchman:2003sq}) to re-expresses exactly general relativity in terms of tetrad fields, $\T{A}{^\ha _\mu}$, which play the role of vector potentials in Maxwell's equations.
As elucidated in the case of binary black hole collisions, general relativity expressed as a nonlinear electrodynamics system gives rise to a rich phenomenology of plasma-like dynamics \cite{Boyeneni:2025tsx}.

As a starting point, we introduce
\begin{align}
    \T{A}{^\mu_\ha} \T{A}{^\nu_\hb} g_{\mu\nu} = \eta_{\hat{\alpha}\hat{\beta}}\,,
\end{align}
where $\eta_{\ha\hb}$ is the flat spacetime Minkowski metric, and hatted indices are relative to the tetrad observer.
The tetrad fields naturally bring about a Maxwell-like field strength tensor \cite{Olivares_2022}: 
\begin{align}
    \T{F}{^{\hat{\alpha}}_{\mu\nu}} = \partial_\mu \T{A}{^\ha _\nu} - \partial_\nu \T{A}{^\ha _\mu}\,.
\end{align}
This field strength tensor is interpreted as defining a spin connection, $\omega_{\hat{\alpha}\hat{\beta}\hat{\mu}}$. In order to obtain general relativity, this spin connection has to be torsion-free and take Levi-Civita form,
$\omega_{\hat{\alpha}\hat{\beta}\hat{\mu}}
= -\frac{1}{2} \left( F_{\hat{\beta}\hat{\alpha}  \hat{\mu}} + F_{\hat{\mu}\hat{\alpha}  \hat{\beta}} - F_{\hat{\alpha}\hat{\beta}\hat{\mu}}\right)\,,$ where $F_{\hat{\alpha}\hat{\beta}\hat{\mu}}=A^\beta{}_{\hat{\beta}}\T{A}{^\mu_{\hat{\mu}}}F_{\hat{\alpha}\beta\mu}$.
Further, it naturally engenders electric and magnetic fields
\begin{align}
    e^{\ha \hb} &= F^{\ha \hmu \hb} u_{\hmu}\,,\\
    b^{\ha \hb} &= \star F^{\ha \hmu \hb} u_{\hmu}\,,
\end{align}
where $u_\mu$ is a suitable observer, which, for our case, we will define in the next Section.

Accordingly, general relativity arises as a nonlinear version of electrodynamics in a medium where, like a polarization tensor, $\T{F}{^\ha_{\mu\nu}}$ is related to an in-medium tensor 
\begin{align}
    \star \, \T{\cU}{_\ha ^{\hb \hmu}} = \omega^{[\hat{\beta}\hat{\mu}]}{}_{\hat{\alpha}} + \delta^{\hat{\beta}}{}_{\hat{\alpha}} \omega^{[\hat{\mu}\hn]}{}_{\hn} - \delta^{\hat{\mu}}{}_{\hat{\alpha}} \omega^{[\hat{\beta}\hn]}{}_{\hn}
\end{align}
 \cite{Olivares_2022} via the Nester-Witten form \cite{Frauendiener:2006dx}. The Einstein equations in vacuum imply that \cite{Olivares_2022} 
\begin{align}
    0=G^\mu{}_{\alpha} = \frac{1}{\sqrt{-g}} \partial_\nu \left(\sqrt{-g}\, \star \mathcal{U}_{\alpha}{}^{\mu\nu} \right) - t^\mu{}_{\alpha} \, ,
\end{align}
where $G_{\mu\nu}$ is the Einstein tensor, and \cite{Olivares_2022} 
\begin{align}\label{eqn:tmunu_LL}
    t^{\hat{\mu}}{}_{\hat{\nu}} = F^{\hat{\alpha}}{}_{\hat{\beta}\hat{\nu}}\star \mathcal{U}_{\hat{\alpha}}{}^{\hat{\beta}\hat{\mu}} - \frac{1}{4} \delta^{\hat{\mu}}{}_{\hat{\nu}} F^{\hat{\alpha}}{}_{\hat{\beta}\hat{\lambda}}\star \mathcal{U}_{ \hat{\alpha}}{}^{\hat{\beta}\hat{\lambda}}
\end{align}
is the energy--momentum pseudo-tensor of spacetime. Although not exactly identical, this equation resembles the conservation of the Landau-Lifshitz pseudo-tensor \cite{landau1975classical,Clough:2021qlv}. 
Because of the anti-symmetry of $\T{\cU}{_\ha ^{\mu\nu}}$, we naturally have an expression for energy--momentum conservation of a vacuum spacetime \cite{Olivares_2022}:
\begin{align}\label{eqn:LL}
    \partial_\mu\left( \sqrt{-g}\, \T{t}{^{\mu}_{\hat{\alpha}}} \right)=0\,.
\end{align}
This equation forms the starting point of our analysis, analogous to the conservation of a matter stress--energy tensor in Eq. \eqref{eqn:tmunu_MHD}.

\section{Elsasser formulation of general relativity}
\label{sec:elsasser}
In order to demonstrate that general relativity possesses an Elsasser equation and, subsequently, an Alfv\'enic nonlinear interaction structure, we must show that (under the proper assumptions) Eq. \eqref{eqn:LL} takes a form comparable to that of Eq. \eqref{eqn:elsasser_mhd}.
To facilitate this process, we borrow an important insight and analogy from force-free electrodynamics \cite{Thompson:1998ss,Gruzinov:1999aza}. In that system, the energy--momentum tensor is essentially Maxwell-like. 
However, a preferred frame exists in which the Lorentz force vanishes; i.e., the system is force-free. It has been shown both mathematically \cite{Thompson:1998ss,TenBarge:2021qmk} and numerically \cite{Cho:2004nn,Zrake:2015hda,Li:2018kco,Ripperda:2021pzt} that relativistic force-free electrodynamics exhibits Alfv\'enic turbulence. 
Mathematically, the frame requirement in force-free electrodynamics posits that there is a velocity, $u_{\rm FF}^\mu$, such that the electromagnetic Lorentz force, $e_{\rm EM}^\nu = u_\mu^{\rm FF} F^{\nu\mu}_{\rm EM}=0$, vanishes.
In the electromagnetic reformulation of general relativity, the geodesic force is similarly given by the comoving electric field $e^{\ha \hb}$, meaning ${\rm d} u^{\ha}/{\rm d}\hat{t} = e^{\hb\ha}u_{\hb}$. 
In analogy with the force-free electrodynamics case, we choose a freely-falling tetrad.
Following Ref. \cite{Maluf:2023rwe}, the evolution of the tetrad generally follows
\begin{align}
    \frac{D \T{A}{^\ha_\mu}}{Ds} = 
    \T{\phi}{^\ha_\hb}\T{A}{^\hb_\mu}\,,
\end{align}
where $\phi^{{\ha}{\hb}}=\T{A}{^\mu_{\hat{0}}} \T{\omega}{^{\ha\hb}_\mu}$ encodes rotations ($\phi^{\hat{i}\hat{j}}$) and accelerations ($\phi^{\hat{\alpha}\hat{0}}$), and $s$ is the coordinate along the tetrad wordline. 
Here, we choose to align the tetrad with the worldline of the force-free observer $\T{A}{^\mu _{\hat{0}}}=u^\mu$.
Fermi-Walker transport of the tetrad, i.e., neglecting artificial frame rotations, requires that  
$\T{\phi}{^{\ha}_{\hb}} = {\T{e}{^{\ha}_{\hb}} - \T{e}{_{\hb}^{\ha}}} =0$ \cite{Maluf:2023rwe}.
As a consequence, we find that 
$0 = u_{\hn}\T{F}{^\hn _{\ha\hb}} = - \varepsilon_{\ha \hb\hk\hmu} u^{\hk} u_{\hn} b^{\hn \hmu}$ and $e_{\ha\hb}=e_{\hb\ha}$.
Together, these imply the following conditions for the freely-falling electric and magnetic fields:{
\begin{align}\label{eqn:freefall}
    e^{\ha \hb} u_{\ha}=0 ~,~ u_{\ha} b^{\ha\hb} = 0\,.
\end{align}
This is in full analogy to force-free electrodynamics.
}

However, just demanding Eq. \eqref{eqn:freefall} is not sufficient for a description of turbulence; everything we have done so far holds for any spacetime. The assumption of a turbulent cascade implies energy transfer from large to small scales. On sufficiently small scales, $\ell$, we assume the background net curvature is negligible such that the freely-falling frame of $u^\mu$ at a given point is also a good approximation in a local neighborhood. In other words, consider a perturbation $u^\mu \rightarrow u^\mu +\delta u^\mu$ at a neighboring point separated by the distance $\ell^{\hn}$. Starting from the Riemann tensor $R_{\ha\hb\hmu\hn}$, geodesic deviation then implies that the relative acceleration
$ \Delta a_{\ha} = R_{\ha\hb\hmu\hn} \delta u^{\hb} \delta u^{\hmu} \ell^{\hn} = u^\mu \nabla_\mu \left[e^{\ha\hb} \ell_{\ha}\delta u_{\hb} \right] +\mathcal{O}(\delta u^3)$ should be small (i.e., $e^{\ha\hb}\ell_{\ha}\ll 1$), since this should hold for any scale $\ell < l_{\rm curv}$,  where $l_{\rm curv} \simeq 1/{{\rm R}\, \ell}$ is the local curvature scale. {In essence, we consider the scale at which the tangent space is a good approximation of the local curvature.}
Similar approximations are common in MHD turbulence as well, where the scale of the local cascade is assumed to be smaller than the inertial scale. Therefore, we posit that we can (at least locally) assume the following, stronger, condition on the background metric:
\begin{align}\label{eqn:e}
    e^{\ha\hb}\approx0\,,
\end{align}
where $e^{\ha\hb}$ is still expressed in the frame of a freely-falling observer.

As a consequence of this freely-falling observer \eqref{eqn:freefall} with a suitable locally uniform background curvature \eqref{eqn:e}, we find that the Maxwell-like field structure takes a simple form,
\begin{align}
    b^{\ha \hb} &= \frac{1}{2} \ve ^{\hmu \hb \hn \hs} u_{\hmu} \T{F}{^\ha _{\hn \hs}}\,,\\
    \label{eqn:Fabc}
    \T{F}{^\ha _{\hb \hmu}} &= - \ve _{\hb \hmu \hn \hs} u^{\hn} b^{\ha \hs}\,,\\
    \star F^{\ha \hb \hmu} &= u^{\hb} b^{\ha \hmu} - u^{\hmu} b^{\ha \hb}\,,
\end{align}
in close analogy with that of ideal MHD.
We can now use these expressions of the field strength tensors in terms of the local gravitational magnetic field to compute the stress--energy tensor \eqref{eqn:tmunu_LL}.
In particular, we have
\begin{align}
    \star \, \T{\cU}{_\ha^{\hb \hn}} 
    &= -\frac{1}{2} \left( 
    \T{F}{_\ha ^{\hb \hn}} 
    + \T{F}{^{\hn \hb} _\ha} - \T{F}{^{\hb \hn} _\ha}
    \right)\nonumber\\
    &\phantom{=} - \T{\delta}{^\hb _\ha} 
    \T{F}{_\hmu ^{\hn \hmu}}
    + \T{\delta}{^\hn _\ha} 
    \T{F}{_\hmu ^{\hb \hmu}} \, ,
\end{align}
yielding
\begin{align}
    \T{F}{^\ha _{\hb \hn}} \star \T{\cU}{_\ha ^{\hb \hn}} 
    &= 2 \T{F}{^\ha _{\hn \ha}} \T{F}{_\hmu ^{\hn \hmu}}
    - \frac{1}{2} \T{F}{^\ha _{\hb \hn}} \T{F}{_\ha ^{\hb \hn}}
    - \T{F}{^\ha _{\hb \hn}} \T{F}{^{\hn \hb}_\ha}\,,\\
    \notag
\T{F}{^\ha _{\hb \hn}} \star \T{\cU}{_\ha ^{\hb \hmu}} 
&= \T{F}{^\ha _{\hn \ha}} \T{F}{_\hb ^{\hmu \hb}}
+ \T{F}{^\hmu _{\hb \hn}} \T{F}{_\ha ^{{\hb} \ha}}\\
&-\frac{1}{2} \left( 
\T{F}{^\ha _{\hb \hn}} \T{F}{_\ha ^{\hb \hmu}}
+ \T{F}{^\ha _{\hb \hn}} \T{F}{^{\hmu \hb}_\ha}
- \T{F}{^\ha _{\hb \hn}} \T{F}{^{\hb \hmu}_\ha}
\right)\,.
\end{align}
Using equation \eqref{eqn:Fabc}, we list some intermediate quantities:
\begin{flalign}
    \T{F}{^\ha _{\hb \hn}} \T{F}{_\ha ^{\hb \hn}} &= 2b^2\,,\\
    \T{F}{^\ha _{\hb \hn}} \T{F}{^{\hn \hb}_\ha} &=
    b^2 - \Tr(b)^2\,,\\
    \T{F}{^\ha _{\ha \hn}} \T{F}{_\hmu ^{\hn \hmu}} &=
    - b^2 + b^{\ha \hb} b_{\hb \ha}\,,\\
    \notag
    \T{F}{^\ha _{\ha \hn}} \T{F}{_\hb ^{\hmu \hb}} &=
    - \T{\delta}{^\hmu _\hn} (b^2 - b^{\ha \hb} b_{\hb \ha}) - u^{\hmu}u_{\hn} (b^2 - b^{\ha \hb}b_{\hb \ha})\\
     &+ b^{\hmu \ha} b_{\hn \ha} - b^{\hmu \ha} b_{\ha \hn}
    + b^{\ha \hmu} b_{\ha \hn} - b^{\ha \hmu} b_{\hn \ha}\,,\\
    \T{F}{^\hmu _{\hb \hn}} \T{F}{_\ha ^{\hb \ha}} &=
    b^{\hmu \ha} b_{\hn \ha} - b^{\hmu \ha} b_{\ha \hn}\,,\\
    \T{F}{^\ha _{\hb \hn}} \T{F}{_\ha ^{\hb \hmu}} &=
    \T{\delta}{^{\hmu}_{\hn}} b^2 - b^{\ha \hmu}b_{\ha \hn}
    + u^{\hmu}u_{\hn} b^2\,,\\
    \T{F}{^\ha _{\hb \hn}} \T{F}{^{\hmu \hb}_\ha} &=
    b^{\hmu \ha}b_{\hn \ha} - \Tr(b) \T{b}{^\hmu _\hn}\,,\\
    \notag
    \T{F}{^\ha _{\hb \hn}} \T{F}{^{\hb \hmu}_\ha} &=
    - \T{\delta}{^\hmu _\hn} (b^2 - \Tr(b)^2) - u^{\hmu}u_{\hn} (b^2 - \Tr(b)^2)\\
     &+ b^{\hmu \ha} b_{\hn \ha} - \Tr(b) \T{b}{^\hmu _\hn}
    + b^{\ha \hmu} b_{\ha \hn} - \Tr(b) \T{b}{_\hn ^\hmu}\,,
\end{flalign}
where we defined $\Tr(b) = \T{b}{^{\ha}_{\ha}}$ and \(b^2 = b^{\ha \hb} b_{\ha \hb}\).
Putting everything together into Eq. \eqref{eqn:tmunu_LL}, we obtain the concise expression
\begin{align}\label{eqn:tmunu_LL_gr}
\notag
\T{t}{^\hmu_\hn} &= u^{\hmu} u_{\hn} \left( 
 \frac{\Tr(b)^2}{2} - b^{\ha \hb}b_{\hb \ha}\right)
+ \frac{\T{\delta}{^\hmu _\hn}}{2} \left( 
 \frac{\Tr(b)^2}{2} - b^{\ha \hb}b_{\hb \ha}\right)\\
&- \frac{\Tr(b)}{2} \T{b}{_\hn^\hmu} +
b^{\ha \hmu} b_{\hn \ha} \,.
\end{align}
This tensor bears remarkable resemblence to the stress--energy tensor of ideal MHD and force-free electrodynamics equation \eqref{eqn:tmunu_MHD}.

Parallel to the MHD case \cite{Chandran:2017zdg}, we define gravitational Elsasser variables

\begin{flalign}\label{eqn:z_gr}
    \T{z}{^\hb _{\ha}}_\pm &= u^{\hb} u_{\ha} \pm \frac{{^{\rm TF \,}\!}\T{b}{_\ha ^\hb}}{\sqrt{\mathcal{E}}}\,,
\end{flalign}
as well as 
\begin{flalign}
{^{\rm TF \,}\!}\T{b}{_\ha ^\hb}&=
\T{b}{_\ha ^\hb} - \frac{1}{4}\Tr(b)\T{\delta}{^\hb _\ha}\,,\\
    \mathcal{E} &= \frac{\Tr(b)^2}{2} - b^{\ha \hb}b_{\hb \ha}\,,\\
    \Pi &= \frac{\Tr(b)^2}{16\mathcal{E}} - \frac{1}{2}\,,
\end{flalign}
where we have introduced the trace free gravitational magnetic field tensor $ {^{\rm TF \,}\!}\T{b}{_\ha ^\hb}$.
These variables strongly resemble their MHD counterparts in Eq. \eqref{eqn:z_mhd}.
However, because of the tensorial nature of gravitational waves, now the Elsasser variables need to be rank-2 tensors.

At this point, we are in a position to obtain a gravitational wave Elsasser equation by combining Eqs. \eqref{eqn:tmunu_LL_gr} and \eqref{eqn:z_gr}. In particular, we find 
\begin{align}
    \notag
    \T{z}{^\hmu _\ha}_\pm \T{z}{^\ha _\hn}_\mp &=
    -u^{\hmu}u_{\hn} - \frac{b^{\ha \hmu}b_{\hn \ha}}{\cE}\\
        \notag
    &+ \frac{\Tr(b)}{2\cE} \T{b}{_\hn ^\hmu} 
    -\frac{\Tr(b)^2}{16\cE} \T{\delta}{^\hmu _\hn} \\
    &= \T{z}{^\hmu _\ha}_\mp \T{z}{^\ha _\hn}_\pm 
\end{align}
and, hence,
\begin{align}
    \T{z}{^\hmu _{\ha}}_\pm \T{z}{^\ha _\hn}_\mp + \Pi \T{\delta}{^\hmu _\hn} &= {-\frac{\T{t}{^\hmu _\hn}}{\cE}}.
\end{align}
Notice that \(\T{z}{^\hmu _{\ha}}_\pm \T{z}{^\ha _\hn}_\mp\) differs from its MHD counterpart in yet another way: its overall sign. In the MHD case, 
\(z^{\hmu} _\pm z^{\hn} _\mp \sim u^{\hmu} u^{\hn}\), whereas  
\(\T{z}{^\hmu _{\ha}}_\pm \T{z}{^\ha _\hn}_\mp
\sim -u^{\hmu} u_{\hn}\) here.

Taking the divergence and enforcing conservation via \eqref{eqn:LL}, at once we obtain an Elsasser equation of general relativity,
\begin{align}\label{eqn:elsasser_gr}
    0=& \partial _\mu \left[{\sqrt{-g}}\left(\T{z}{^\mu _{\ha}}_\pm \T{z}{^\ha _\hn}_\mp + \Pi \T{\delta}{^\mu _\hn}\right)\right]\nonumber \\
    &+ {\sqrt{-g}}\left( {\T{z}{^\mu _\ha}_\pm \T{z}{^\ha _\hn}_\mp}
    + \Pi \T{\delta}{^\mu _\hn}
    \right)\frac{\partial _\mu \mathcal{E}}{\mathcal{E}}.
\end{align}
This equation bears remarkable resemblance to the relativistic MHD Elsasser equation \eqref{eqn:elsasser_mhd} in the force-free electrodynamics limit (where $\mathcal{E} \rightarrow b^2$ and $\Pi \rightarrow 1/2$). Additionally, we observe that the equation allows for couplings to pressure perturbations (i.e., fast magnetosonic modes), and we observe that the fundamental nonlinear interaction will, just as in MHD, arise from couplings between oppositely directed Elsasser variables, $\T{z}{^\mu _{\ha}}_\pm$.

\section{Nonlinear Alfv\'enic interactions}

In MHD, the existence of the Elsasser equation immediately implies a number of important statements about nonlinear interactions of counter-propagating Alfv\'en waves \cite{1965PhFl....8.1385K}, an anisotropic turbulent energy cascade \cite{1995ApJ...438..763G,1996ApJ...465..845N}, and a distinction between the weak and strong Alfv\'enic turbulence regimes, depending on the concept of critical balance \cite{1995ApJ...438..763G}.
To illustrate that these concepts can, in principle, be carried over to general relativity, it must be shown that the fundamental nonlinear character of the Elsasser equation is also preserved for Eq. \eqref{eqn:elsasser_gr}.

Commonly, this is done by perturbing the Elsasser equation around a uniform background (see, e.g., Refs. \cite{Chandran:2017zdg,TenBarge:2021qmk,Mallet:2021fwv}, which we follow here). In reality, the details of this background field may depend on the context in which turbulence is considered \cite{Galtier:2017mve,Ma:2025rnv}.

For simplicity, mirroring a reduced MHD approach \cite{1987JGR....92.7363M}, we  introduce perturbations
\begin{align}
\T{{z}}{^\mu_{\hat{\alpha}_\pm}}  = \T{\bar{z}}{^\mu_{\hat{\alpha}_\pm}} + \delta \T{z}{^\mu_{\hat{\alpha}_\pm}} ~, ~ \Pi = \bar{\Pi} + \delta\Pi ~,~ \mathcal{E} = \bar{\mathcal{E}} + \delta \mathcal{E}\,,
\end{align}
where barred quantities represent averages in the averaged fluid rest frame sense (see discussion around Eq. \eqref{eqn:e} and Ref. \cite{Chandran:2017zdg}). For now, we neglect changes in the volume factor $\sqrt{-g}$, though these may be important, especially in strongly nonlinear cases \cite{Ma:2025rnv}.

Next, we obtain the fluctuating principle part of the equation,
\begin{align}
     \partial_\mu\left[\sqrt{-g}\right. & \left. \left(\T{z}{^\mu _{\ha}}_\pm \T{z}{^{\ha} _\hn}_\mp + \Pi \T{\delta}{^\mu _\hn}\right)\right] \nonumber \\ 
     \notag
     &\phantom{=}- 
    \partial_\mu\left[ \sqrt{-g}\left(\T{\bar{z}}{^\mu _{\ha}}_\pm \T{\bar{z}}{^{\ha} _\hn}_\mp + \bar{\Pi} \T{\delta}{^\mu _\hn}\right)\right]\\
    &\approx \partial_\mu\left[\sqrt{-g}\left(\delta\T{z}{^\mu _{\ha}}_\pm \T{\bar{z}}{^{\ha} _\hn}_\mp + \T{\bar{z}}{^\mu _{\ha}}_\pm \delta\T{{z}}{^{\ha} _\hn}_\mp \right)\right] \nonumber \\ 
    &\phantom{=} + 
    \partial_\mu\left[ \sqrt{-g}\left(\delta\T{{z}}{^\mu _{\ha}}_\pm \delta\T{{z}}{^{\ha} _\hn}_\mp + {\delta \Pi} \T{\delta}{^\mu _\hn}\right)\right]\,.
\end{align}

We introduce correlation lengths $\ell_\perp$ and $\ell_\parallel$, which are taken to be relative to the local magnetic background field ${^{\rm TF \,}}\!\bar{b}^{\hb}_{\ha}$, or, for our purposes, $\T{\bar{z}}{^\mu _{\ha}}_\pm $. It is an important feature of MHD turbulence that the effective fluid variables will exhibit different scalings along and across the field.
This means that in MHD turbulence, gradients along the magnetic field will be on the order of $\mathcal{O}(\ell_\parallel/\epsilon)$ instead of $\mathcal{O}(\epsilon)$, where $\epsilon$ is the effective order parameter. It follows that, unlike terms such as $\delta\T{z}{^\mu_\ha}_\mp \delta \T{z}{^\ha _\hn}_\pm$ that are $\mathcal{O}(\epsilon^2)$, terms like \cite{Chandran:2017zdg}
\begin{align}
    \bar{z}^\mu_{\mp\hb} \partial_\mu \delta {z}^{\hb}_{\ha\pm} \sim  
\delta{z}^\mu_{\mp\hb} \partial_\mu \delta{z}^{\hb}_{\ha\pm}\sim \mathcal{O}(\epsilon)\,.
\end{align}

Consequently, to leading order (see Ref. \cite{Chandran:2017zdg} for the MHD case),
\begin{align}
    0&=\partial_\mu \left[\sqrt{-g}\, \delta \T{t}{^\mu_{\hn}}\right]\\
    &\approx \partial_\mu\left[\sqrt{-g}\left(\delta\T{z}{^\mu _{\ha}}_\pm \T{\bar{z}}{^{{\ha}} _\hn}_\mp + \T{\bar{z}}{^\mu _{\ha}}_\pm \delta\T{z}{^{{\ha}} _\hn}_\mp \right)\right] + N^\pm_{\hat{\nu}} \nonumber\\
    &\phantom{=}+ 
    \sqrt{-g}\left[ \delta \T{z}{^\mu _{\hb}}_\pm \T{\bar{z}}{^{\hb} _{\hn}}_\mp  +  \T{\bar{z}}{^\mu _{\hb}}_\pm \delta\T{z}{^{\hb} _{\hn}}_\mp\right]\partial_\mu \log \bar{\mathcal{E}}\,.
\end{align}

Here, we have introduced the nonlinear interaction term
\begin{align}
    N^\pm_{\hat{\nu}} = 
    \partial_\mu\left[ \sqrt{-g}\left(\delta\T{{z}}{^\mu _{\hb}}_\pm \delta\T{{z}}{^{\hb} _\hn}_\mp + {\delta \Pi} \T{\delta}{^\mu _\hn}\right)\right]\,,
\end{align}
which encodes nonlinear interactions. In particular, the first term is only nonzero if both plus and minus perturbations are present; i.e., the interaction is Alfv\'enic. Furthermore, unlike the Elsasser variables in the MHD case, the tensorial nature of the general relativistic Elsasser variables implies that the polarization content has to match up between those oppositely directed variables. 
This is consistent with findings in the relativistic MHD case \cite{Ripperda:2021pzt}.

We can gain further insight into this equation when considering additional simplifying assumptions in line with force-free electrodynamics \cite{TenBarge:2021qmk}.
First, we assume corrections and fluctuations to $\Pi \simeq \frac{1}{2}$ are small, and $\delta \Pi\simeq + \mathcal{O}(\epsilon^2)$ (this statement is exact in the force-free limit of MHD). Second, relativistic MHD is fundamentally compressible, as the sound speed is bounded by the speed of light. Consequently, reduced MHD in the relativistic case may also admit compressible fast waves. These are normally considered to be small in this context for MHD\cite{Chandran:2017zdg,TenBarge:2021qmk}.
While a full analysis beyond the scope of this work is necessary to render a conclusive verdict on the role compressive modes play in gravitational wave turbulence, we will order out these modes to illustrate the Alfv\'enic structure of the equations.
That is, we assume that $\partial_\mu \T{\bar{z}}{^{\mu}_{\pm\ha}} \sim \partial_\mu \T{\delta \bar{z}}{^{\mu}_{\pm\ha}} \sim \mathcal{O}\left(\epsilon^2\right)$.
Under these stronger assumptions, we obtain, to lowest order,
\begin{align}
\T{\bar{z}}{^\mu_{\ha}}_\mp \partial_\mu \delta \T{z}{^{\ha}_{\hn}}\pm \simeq-& 
\delta \T{z}{^\mu_{\ha}}_\mp \partial_\mu \delta \T{z}{^{\ha}_\hn}_{\pm}  \nonumber\,,\\
 &-\left(\delta \T{z}{^\mu _{\hb}}_\pm \T{\bar{z}}{^{\hb} _{\hn}}_\mp  +  \T{\bar{z}}{^\mu _{\hb}}_\pm \delta\T{z}{^{\hb} _{\hn}}_\mp
\right)\frac{\partial_\mu \bar{\mathcal{E}}}{\bar{\mathcal{E}}}\,.
\end{align}

This equation naturally parallels its MHD counterpart \cite{TenBarge:2021qmk}. We see that the principle part is an advection term of the perturbation, $\delta \T{z}{^{\ha}_{\hn}}_\pm$, along the background state $\T{\bar{z}}{^\mu_{\ha}}_\mp \partial_\mu$, which is altered only by nonlinear couplings between oppositely directed perturbations. This is precisely the Alfv\'enic nature of the interaction.

\section{Conclusions}

Gravitational waves can be intrinsically nonlinear and have recently been shown to generate turbulence, both theoretically \cite{Galtier:2017mve,Galtier:2021ovg,Gay:2024kay} and numerically \cite{Ma:2025rnv}.
Here, we take an essential step toward uncovering the theory of gravitational wave turbulence and nonlinear interaction, mirroring efforts in MHD turbulence.
By exploiting an exact formulation of general relativity as nonlinear electrodynamics \cite{Olivares_2022,Peshkov:2022cbi,Boyeneni:2025tsx}, we have shown that general relativity admits an Elsasser equation \eqref{eqn:elsasser_gr}---the fundamental equation describing MHD turbulence. This equation describes the interaction of perturbations of the gravitational magnetic field relative to a freely-falling observer. The equations we find bear strong resemblence to the MHD Elsasser equations in the force-free electrodynamics limit \cite{TenBarge:2021qmk,Ripperda:2021pzt}. Moreover, we have shown that Alfv\'enic nonlinear wave interactions of these general relativity Elsasser equations derived here require oppositely directed Elssasser variables, just as in MHD turbulence \cite{Chandran:2017zdg}. 

Overall, our work paves the way for a completely novel investigation of gravitational wave turbulence and nonlinearities by making it amenable to an extensive number of treatments developed in the MHD literature \cite{Schekochihin:2020aqu}. Follow-up work will be devoted towards elucidating fully the nonlinear structure and cascade prediction of these equations, the distinction of a weak and strong turbulence regime (including the potential existence and meaning of critical balance) \cite{1995ApJ...438..763G}, as well as connections to numerical investigations \cite{Galtier:2021ovg,Ma:2025rnv}.

\begin{acknowledgments}
ERM is grateful for insightful discussions with James Beattie, Andrei Beloborodov, Maxim Lyutikov, Mikhail Medvedev, Joonas N\"attil\"a, Anatoly Spitkovsky, and Arno Vanthieghem. HK gratefully acknowledges a summer research fellowship from the Department of Mathematics at Caltech.
ERM acknowledges partial support by the National Science Foundation under grants No. PHY-2309210. This research was supported in part by grant NSF PHY-2309135 to the Kavli Institute for Theoretical Physics (KITP).
\end{acknowledgments}

\bibliography{els}
\end{document}